# Optical Angular Momentum in Classical Electrodynamics

Masud Mansuripur

College of Optical Sciences, The University of Arizona, Tucson

**Abstract**. Invoking Maxwell's classical equations in conjunction with expressions for the electromagnetic (EM) energy, momentum, force, and torque, we use a few simple examples to demonstrate the nature of the EM angular momentum. The energy and the angular momentum of an EM field will be shown to have an intimate relationship; a source radiating EM angular momentum will, of necessity, pick up an equal but opposite amount of mechanical angular momentum; and the spin and orbital angular momenta of the EM field, when absorbed by a small particle, will be seen to elicit different responses from the particle.

**1. Introduction**. The goal of the present paper is to use a few simple examples to illustrate some of the fundamental properties of the electromagnetic (EM) field. Using elementary physical concepts and mathematical methods, we attempt to explain the nature of interactions that involve exchanges of energy, linear momentum, and angular momentum between EM fields and material media. To simplify the analysis, we shall assume that the material media are charge-neutral, solid, non-magnetic, and linearly polarizable in the presence of an external electric field.

It is well known that, in free space, the EM fields $\boldsymbol{E}(\boldsymbol{r},t)$ and $\boldsymbol{H}(\boldsymbol{r},t)$ have energy-density $\mathcal{E}(\boldsymbol{r},t) = ½\varepsilon_0\boldsymbol{E}\cdot\boldsymbol{E} + ½\mu_0\boldsymbol{H}\cdot\boldsymbol{H}$, which propagates at a rate (per unit area per unit time) given by the Poynting vector $\boldsymbol{S}(\boldsymbol{r},t) = \boldsymbol{E}\times\boldsymbol{H}$. These fields also possess EM momentum-density $\boldsymbol{\wp}(\boldsymbol{r},t) = \boldsymbol{S}(\boldsymbol{r},t)/c^2$, and EM angular momentum-density $\boldsymbol{\mathcal{L}}(\boldsymbol{r},t) = \boldsymbol{r}\times\boldsymbol{\wp}(\boldsymbol{r},t)$, the latter being expressed relative to the origin of coordinates ($\boldsymbol{r}=0$). In the above expressions, $\boldsymbol{r} = x\hat{\boldsymbol{x}} + y\hat{\boldsymbol{y}} + z\hat{\boldsymbol{z}}$ and $t$ represent the coordinates of an arbitrary point in spacetime. In the International System of Units (*SI*), $\varepsilon_0 \cong 8.854\times10^{-12}$ farad/m is the permittivity of free space, $\mu_0 = 4\pi\times10^{-7}$ henry/m is the permeability of free space, and $c = 1/\sqrt{\mu_0\varepsilon_0}$ is the speed of light in vacuum [1-3].

It is also well-known that, if the electric-dipole-density of a material medium (commonly referred to as its polarization) is given by $\boldsymbol{P}(\boldsymbol{r},t)$, then the rate of exchange of EM energy-density between the dipoles and the local fields is $\partial\mathcal{E}(\boldsymbol{r},t)/\partial t = \boldsymbol{E}\cdot\partial\boldsymbol{P}/\partial t$, the EM force-density exerted by the fields on the dipoles is $\boldsymbol{F}(\boldsymbol{r},t) = (\boldsymbol{P}\cdot\boldsymbol{\nabla})\boldsymbol{E} + (\partial\boldsymbol{P}/\partial t)\times\mu_0\boldsymbol{H}$, and the EM torque-density experienced by the dipoles is $\boldsymbol{T}(\boldsymbol{r},t) = \boldsymbol{r}\times\boldsymbol{F} + \boldsymbol{P}\times\boldsymbol{E}$. Note that we have deliberately avoided the use of the $B$-field in the above equations. In general, $\boldsymbol{B}(\boldsymbol{r},t) = \mu_0\boldsymbol{H} + \boldsymbol{M}$, where $\boldsymbol{M}(\boldsymbol{r},t)$ represents the magnetization of the material medium. In the absence of magnetization, of course, we will have $\boldsymbol{B} = \mu_0\boldsymbol{H}$. Since the present paper is not concerned with magnetic or magnetizable media, we have ignored the contributions of $\boldsymbol{M}(\boldsymbol{r},t)$ to EM energy, force, and torque in the preceding equations. Suffice it to say that, when magnetization is present, additional terms must be introduced in order to account for the interactions between $\boldsymbol{M}(\boldsymbol{r},t)$ and the local $\boldsymbol{E}$ and $\boldsymbol{H}$ fields [4-7].

Perhaps less well-known are the mechanisms of momentum exchange between EM fields and material media, the relation between EM energy and angular momentum, and certain subtle differences between the spin and orbital angular momenta of the EM field [8-18]. The issues pertaining to light-matter interaction, and the nuances of optical linear and angular momenta, have been the subject of discussion (and sometimes controversy) for over a century; the interested reader can find in the published literature reports of experimental and theoretical investigations as well as excellent reviews of the subject dating as far back as the early years of the twentieth century [19-25]. The present paper is an attempt at clarifying some of the conceptual issues that lie at the heart of the interactions between light and matter involving EM force and torque as well as linear and angular momenta. Some of our examples involve idealized, linearly polarizable particles whose EM properties are encapsulated in an electric-dipole-density function $\boldsymbol{P}(\boldsymbol{r},t)$, which is a differentiable function of space and time. Given that, by definition, $\boldsymbol{P}(\boldsymbol{r},t)$ is a locally-averaged entity, namely, the dipole moment in a finite volume $\Delta v$ divided by $\Delta v$, it is imperative to recall that $\Delta v$ must be large enough to contain a significant number of atoms/molecules in order to validate an "average" description. Thus, when we invoke a point-dipole whose polarization is represented (for the sake of mathematical convenience) by a Dirac delta-function, e.g., $\boldsymbol{P}(\boldsymbol{r},t)=\boldsymbol{p}(t)\delta(\boldsymbol{r})$, the reader should understand that our intent is to describe a spherical particle that is much smaller than the wavelength of the EM field acting on the particle, yet is sufficiently large to contain a significant number of atoms/molecules.

We begin in Sec. 2 by describing the radiation force experienced by a small, polarizable particle which has a predetermined dielectric susceptibility. Section 3 relates the dielectric susceptibility of small spherical particles to their refractive index, while properly accounting for the effects of radiation resistance. The relation between the energy and the angular momentum content of a circularly-polarized plane-wave is the subject of Sec. 4. In Sec. 5 we show that a current sheet emitting radiation that carries optical angular momentum must, of necessity, experience a mechanical torque in order to balance the overall angular momentum of the system. Section 6 is devoted to a description of vector cylindrical harmonics, which are the eigen-modes of Maxwell's equations in a linear, homogeneous, isotropic, cylindrically-symmetric system. Here we derive the relation between the energy and the orbital angular momentum of a cylindrical harmonic EM wave trapped within a hollow, perfectly-electrically-conducting cylinder. Finally, in Sec. 7 we explore the relations among the energy, linear momentum, and angular momentum picked up by a small particle under illumination by a cylindrical harmonic EM wave. In light of this analysis, it becomes clear why a small particle spins around its own axis when illuminated by a light beam that carries spin angular momentum, whereas the same particle tends to orbit around an axis of vorticity when exposed to a beam (such as a vector cylindrical harmonic) that possesses orbital angular momentum. We close the paper with a summary and a few general remarks in Sec. 8.

**2. Electromagnetic force exerted on a polarizable point-particle**. In the vicinity of $\boldsymbol{r}=\boldsymbol{r}_0$, a monochromatic EM wave of frequency $\omega$ has the $E$-field $\boldsymbol{E}(\boldsymbol{r},t)=\mathrm{Re}[\boldsymbol{E}(\boldsymbol{r})\exp(-\mathrm{i}\omega t)]$ and the $H$-field $\boldsymbol{H}(\boldsymbol{r},t)=\mathrm{Re}[\boldsymbol{H}(\boldsymbol{r})\exp(-\mathrm{i}\omega t)]$. Let a small test-particle, whose electric susceptibility is denoted by $\chi(\omega)=|\chi|\exp(\mathrm{i}\varphi)$, sit at $\boldsymbol{r}=\boldsymbol{r}_0$. The particle is assumed to be a homogeneous solid sphere whose radius is so much smaller than the vacuum wavelength $\lambda_0=2\pi c/\omega$ of the EM wave that it can be treated mathematically as a point-dipole. Needless to say, the particle must be large enough to contain a sufficient number of atoms/molecules, so that an average polarization $\boldsymbol{P}(\boldsymbol{r},t)=\mathrm{Re}[\varepsilon_0\chi(\omega)\boldsymbol{E}(\boldsymbol{r})\exp(-\mathrm{i}\omega t)]$ can be associated with its electric-dipole-density in the context of Maxwell's macroscopic equations. The self-field of the oscillating dipole gives rise to its radiation resistance, which is incorporated into the susceptibility $\chi(\omega)$ of the particle, as explained in the following section. The time-averaged force-density exerted by the external EM field on the test-particle is thus found to be

$$\begin{aligned}\langle\boldsymbol{F}(\boldsymbol{r}_0,t)\rangle&=\langle(\boldsymbol{P}\cdot\boldsymbol{\nabla})\boldsymbol{E}+(\partial\boldsymbol{P}/\partial t)\times\mu_0\boldsymbol{H}\rangle\\&=\tfrac{1}{2}\mathrm{Re}\{[\varepsilon_0\chi(\omega)\boldsymbol{E}(\boldsymbol{r})\cdot\boldsymbol{\nabla}]\boldsymbol{E}^*(\boldsymbol{r})-\mathrm{i}\omega\varepsilon_0\chi(\omega)\boldsymbol{E}(\boldsymbol{r})\times\mu_0\boldsymbol{H}^*(\boldsymbol{r})\}_{\boldsymbol{r}=\boldsymbol{r}_0}\\&=\tfrac{1}{2}\varepsilon_0|\chi|\mathrm{Re}\{\exp(\mathrm{i}\varphi)\,[\boldsymbol{E}(\boldsymbol{r})\cdot\boldsymbol{\nabla}]\boldsymbol{E}^*(\boldsymbol{r})+\exp(\mathrm{i}\varphi)\,\boldsymbol{E}(\boldsymbol{r})\times[\boldsymbol{\nabla}\times\boldsymbol{E}^*(\boldsymbol{r})]\}_{\boldsymbol{r}=\boldsymbol{r}_0}\end{aligned}$$

$$= \tfrac{1}{2}\varepsilon_0|\chi|\mathrm{Re}\big\{\exp(\mathrm{i}\varphi)\,\big\{(E_x\partial_x\boldsymbol{E}^* + E_y\partial_y\boldsymbol{E}^* + E_z\partial_z\boldsymbol{E}^*) + (E_x\hat{\boldsymbol{x}} + E_y\hat{\boldsymbol{y}} + E_z\hat{\boldsymbol{z}})$$
$$\times\,[(\partial_y E_z^* - \partial_z E_y^*)\hat{\boldsymbol{x}} + (\partial_z E_x^* - \partial_x E_z^*)\hat{\boldsymbol{y}} + (\partial_x E_y^* - \partial_y E_x^*)\hat{\boldsymbol{z}}]\big\}\big\}$$

$$= \tfrac{1}{2}\varepsilon_0|\chi|\mathrm{Re}\big\{\exp(\mathrm{i}\varphi)\,[(E_x\partial_x E_x^* + \cancel{E_y\partial_y E_x^*} + \cancel{E_z\partial_z E_x^*})\hat{\boldsymbol{x}}$$
$$+(\cancel{E_x\partial_x E_y^*} + E_y\partial_y E_y^* + \cancel{E_z\partial_z E_y^*})\hat{\boldsymbol{y}}$$
$$+(\cancel{E_x\partial_x E_z^*} + \cancel{E_y\partial_y E_z^*} + E_z\partial_z E_z^*)\hat{\boldsymbol{z}}$$
$$+(E_y\partial_x E_y^* - \cancel{E_y\partial_y E_x^*} - \cancel{E_z\partial_z E_x^*} + E_z\partial_x E_z^*)\hat{\boldsymbol{x}}$$
$$+(E_z\partial_y E_z^* - \cancel{E_z\partial_z E_y^*} - \cancel{E_x\partial_x E_y^*} + E_x\partial_y E_x^*)\hat{\boldsymbol{y}}$$
$$+(E_x\partial_z E_x^* - \cancel{E_x\partial_x E_z^*} - \cancel{E_y\partial_y E_z^*} + E_y\partial_z E_y^*)]\hat{\boldsymbol{z}}\big\}$$

The step-by-step derivations in red color are included here (and also in the following sections) in case the readers would like to check the validity of the results. These lines have been removed from the published version of the article in *Physica Scripta*.

$$= \tfrac{1}{2}\varepsilon_0|\chi|\mathrm{Re}\big\{\exp(\mathrm{i}\varphi)\,\big[(E_x\partial_x E_x^* + E_y\partial_x E_y^* + E_z\partial_x E_z^*)\hat{\boldsymbol{x}}$$
$$+(E_x\partial_y E_x^* + E_y\partial_y E_y^* + E_z\partial_y E_z^*)\hat{\boldsymbol{y}} + (E_x\partial_z E_x^* + E_y\partial_z E_y^* + E_z\partial_z E_z^*)\hat{\boldsymbol{z}}\big]\big\}$$

$$= \tfrac{1}{4}\varepsilon_0|\chi|\cos\varphi\,\boldsymbol{\nabla}(\boldsymbol{E}\cdot\boldsymbol{E}^*) + \tfrac{1}{2}\varepsilon_0|\chi|\sin\varphi\,\mathrm{Im}\big(E_x^*\boldsymbol{\nabla}E_x + E_y^*\boldsymbol{\nabla}E_y + E_z^*\boldsymbol{\nabla}E_z\big). \tag{1}$$

The first term on the right-hand-side of the above equation, being proportional to the gradient of the $E$-field intensity at the location of the test-particle, yields the density of the so-called gradient force exerted by the EM field on the induced dipole. The gradient force is seen to be proportional to the real part $|\chi|\cos\varphi$ of the electric susceptibility. The remaining term on the right-hand-side of Eq.(1), known as the density of the scattering force, is proportional to the imaginary part $|\chi|\sin\varphi$ of the susceptibility.

In the case of a simple plane-wave, we have $\boldsymbol{E}(\boldsymbol{r}) = \boldsymbol{E}_0\exp(\mathrm{i}\boldsymbol{k}\cdot\boldsymbol{r})$. Here $\boldsymbol{E}_0 = \boldsymbol{E}_0' + \mathrm{i}\boldsymbol{E}_0''$ is the complex $E$-field amplitude, and $\boldsymbol{k} = (\omega/c)\hat{\boldsymbol{k}}$ is the real-valued propagation vector in free space. (The unit-vector $\hat{\boldsymbol{k}}$ specifies the direction of propagation.) It is readily verified that the gradient force in this case vanishes, while the scattering force becomes $\tfrac{1}{2}\varepsilon_0|\chi|\sin\varphi\,(\omega/c)(\boldsymbol{E}_0\cdot\boldsymbol{E}_0^*)\hat{\boldsymbol{k}}$.

These results form a foundation for our subsequent discussion pertaining to optical angular momentum. Before addressing the issues related to angular momentum, however, we need to explain the relation between a point-particle's susceptibility and its refractive index.

**3. Relation of particle susceptibility to refractive index and radiation resistance**. Assuming that the $E$-field at $\boldsymbol{r} = \boldsymbol{r}_0$ is linearly-polarized, with amplitude $\boldsymbol{E}(\boldsymbol{r}_0, t) = E_0\cos(\omega t)\,\hat{\boldsymbol{x}}$, and that the test-particle has volume $v$, its dipole moment will be $\boldsymbol{p}(t) = \varepsilon_0|\chi|vE_0\cos(\omega t - \varphi)\,\hat{\boldsymbol{x}}$. As pointed out previously, the implicit assumption here is that the radius of the spherical particle is well below the wavelength $\lambda_0 = 2\pi c/\omega$ of the EM field, so that the local value of the $E$-field as well as its gradient are all that one needs in order to describe the interaction between the EM field and the test-particle, yet the particle's volume $v$ is sufficiently large to allow the association of a macroscopic polarization $\boldsymbol{P}(\boldsymbol{r}, t)$ with the material medium of the particle. The time-averaged rate at which the EM field imparts energy to the dipole will then be

$$\langle \boldsymbol{E}(\boldsymbol{r}_0, t)\cdot\partial\boldsymbol{p}/\partial t\rangle = \tfrac{1}{2}\varepsilon_0|\chi|v\omega E_0^2\sin\varphi. \tag{2}$$

If the test-particle does not absorb any of the energy, all the energy imparted by the field must be re-radiated. It is well-known that the point-dipole $p_0\cos(\omega t)\,\hat{\boldsymbol{x}}$ radiates energy at the time-averaged rate of $\mu_0 p_0^2 \omega^4/(12\pi c)$ [2,3]. Consequently, the relation $|\chi|/\sin\varphi = 3\lambda_0^3/(4\pi^2 v)$, where $\lambda_0 = 2\pi c/\omega$ is the vacuum wavelength of the light, must be satisfied.

In general, the refractive index of a material medium (measured in its bulk form) is a complex-valued function of frequency, $n(\omega) = \sqrt{\mu(\omega)\varepsilon(\omega)}$, where $\mu$ and $\varepsilon$ are the relative permeability and permittivity of the medium [1-3]. For non-magnetic media, $\mu(\omega) = 1$ and the bulk electric susceptibility is given by $\chi_{\text{bulk}}(\omega) = \varepsilon(\omega) - 1 = n^2 - 1$. When a small spherical particle of volume $v$ ($\ll \lambda_0^3$), bulk susceptibility $\chi_{\text{bulk}}(\omega)$, and susceptibility $\chi(\omega)$ to the applied field, is subjected to an external $E$-field, $E_0 \exp(-\mathrm{i}\omega t)\,\hat{\boldsymbol{x}}$, the self-field [26,27] produced by the polarization $\boldsymbol{P}(t) = \varepsilon_0 \chi(\omega) E_0 \exp(-\mathrm{i}\omega t)\,\hat{\boldsymbol{x}}$ of the oscillating dipole will be

$$\boldsymbol{E}_{\text{self}}(t) = -\,[1 - \mathrm{i}(4\pi^2 v/\lambda_0^3)]\boldsymbol{P}(t)/(3\varepsilon_0). \tag{3}$$

Consequently,

$$\varepsilon_0 \chi(\omega) E_0 = \varepsilon_0 \chi_{\text{bulk}}(\omega)\{1 - ⅓[1 - \mathrm{i}(4\pi^2 v/\lambda_0^3)]\chi(\omega)\}E_0. \tag{4}$$

The preceding equation may be solved for the susceptibility $\chi(\omega) = |\chi| \exp(\mathrm{i}\varphi)$ of the point-particle to the applied $E$-field, yielding

$$\chi(\omega) = \frac{3}{[1 + (3/\chi_{\text{bulk}})] - \mathrm{i}(4\pi^2 v/\lambda_0^3)} = \frac{3}{[(n^2+2)/(n^2-1)] - \mathrm{i}(4\pi^2 v/\lambda_0^3)}. \tag{5}$$

Thus, a knowledge of the (generally complex-valued) refractive index $n(\omega)$ of the particle's host material (as measured in the bulk) yields, via Eq.(5), the susceptibility to the external $E$-field, $\chi(\omega)$, of the (spherical) point-dipole of volume $v$.

In the bulk material, where radiation resistance does *not* act as an energy loss mechanism, the second term in the denominator of Eq.(5) is absent, yielding the well-known expression for the *intrinsic* susceptibility (also known as the Clausius-Mossotti relation or the Lorenz-Lorentz formula [1,2]), as follows:

$$\chi_{\text{int}}(\omega) = \frac{3\chi_{\text{bulk}}}{\chi_{\text{bulk}} + 3} = \frac{3(n^2 - 1)}{n^2 + 2}. \tag{6}$$

The intrinsic susceptibility of a material medium thus differs from its bulk susceptibility due to a local field correction that discounts the action of the local $E$-field, $\boldsymbol{P}(\boldsymbol{r}, t)/(3\varepsilon_0)$, produced by a (spherical) dipole upon itself. The intrinsic susceptibility is also what one obtains from the Lorentz oscillator model [1], namely, $\chi_{\text{int}}(\omega) = \omega_p^2/(\omega_0^2 - \omega^2 - \mathrm{i}\gamma\omega)$, which is based on the mass-and-spring model of an atomic system having resonance frequency $\omega_0$, damping coefficient $\gamma$, and plasma frequency $\omega_p$. Substituting the Lorentz oscillator model's expression for $\chi_{\text{int}}(\omega)$ in Eq.(5), we find

$$\chi(\omega) = \frac{\omega_p^2}{\omega_0^2 - \omega^2 - \mathrm{i}[\gamma\omega + (4\pi^2\omega_p^2/3)(v/\lambda_0^3)]}. \tag{7}$$

Clearly, in the case of a small spherical test-particle, the intrinsic damping coefficient $\gamma$ of the dipole's host medium is augmented by the effects of radiation resistance.

**4. Spin angular momentum of circularly polarized plane-wave**. Circularly polarized plane-waves carry EM angular momentum [8-12]. A simple way to demonstrate this involves monitoring the torque exerted by the EM field on a small particle placed in the beam's path. A circularly-polarized plane-wave propagating along the $z$-axis has the following $\boldsymbol{E}$ and $\boldsymbol{H}$ fields:

$$\boldsymbol{E}(\boldsymbol{r},t) = E_0[\sin(k_0 z - \omega t)\,\hat{\boldsymbol{x}} - \cos(k_0 z - \omega t)\,\hat{\boldsymbol{y}}], \tag{8a}$$

$$\boldsymbol{H}(\boldsymbol{r},t) = (E_0/Z_0)[\cos(k_0 z - \omega t)\,\hat{\boldsymbol{x}} + \sin(k_0 z - \omega t)\,\hat{\boldsymbol{y}}]. \tag{8b}$$

In the above equations, $k_0 = \omega/c$ is the wavenumber, and $Z_0 = \sqrt{\mu_0/\varepsilon_0} \cong 377\Omega$ is the impedance of free space. By definition, the field described by Eq.(8) is right-circularly polarized (RCP) because its $E$-field rotates clockwise when seen by an observer facing the source. Let a small particle located at an arbitrary point $(x_0, y_0, z_0)$ have dielectric susceptibility $\varepsilon_0\chi(\omega) = \varepsilon_0|\chi|\exp(\mathrm{i}\varphi)$. The relative susceptibility $\chi(\omega)$ is dimensionless, while its phase angle $\varphi$, being in the interval $(0,\pi)$, indicates that the dielectric (or metallic) particle is absorptive. The material polarization induced within the particle by the plane-wave's $E$-field is given by

$$\boldsymbol{P}(\boldsymbol{r},t) = \varepsilon_0|\chi(\omega)|E_0[\sin(k_0 z_0 - \omega t + \varphi)\,\hat{\boldsymbol{x}} - \cos(k_0 z_0 - \omega t + \varphi)\,\hat{\boldsymbol{y}}]. \tag{9}$$

The rate (per unit time per unit volume) at which energy is absorbed by the particle may now be evaluated as follows:

$$\begin{aligned} d\mathcal{E}/dt &= \boldsymbol{E}\cdot(d\boldsymbol{P}/dt) \\ &= \varepsilon_0|\chi|E_0^2\omega[-\sin(k_0 z_0 - \omega t)\cos(k_0 z_0 - \omega t + \varphi) + \cos(k_0 z_0 - \omega t)\sin(k_0 z_0 - \omega t + \varphi)] \\ &= \varepsilon_0|\chi|E_0^2\omega\sin\varphi. \end{aligned} \tag{10}$$

Note that, for energy to be taken away from the incoming beam, it is essential that the susceptibility's phase angle $\varphi$ be in the $(0,\pi)$ interval. The particle may convert the absorbed energy given by Eq.(10) into heat or other forms of energy, or it may simply radiate the energy away (i.e., loss by scattering). Also taken up by the particle are the linear and angular momenta of the absorbed light, which manifest themselves in the force and torque experienced by the particle. The Lorentz force-density acting on the particle is given by

$$\begin{aligned} \boldsymbol{F} &= (\boldsymbol{P}\cdot\boldsymbol{\nabla})\boldsymbol{E} + (\partial\boldsymbol{P}/\partial t)\times\mu_0\boldsymbol{H} \\ &= \varepsilon_0\mu_0|\chi|Z_0^{-1}E_0^2\omega[-\cos(k_0 z_0 - \omega t + \varphi)\sin(k_0 z_0 - \omega t) + \sin(k_0 z_0 - \omega t + \varphi)\cos(k_0 z_0 - \omega t)]\hat{\boldsymbol{z}} \\ &= \varepsilon_0|\chi|E_0^2(\omega/c)\sin\varphi\,\hat{\boldsymbol{z}}. \end{aligned} \tag{11}$$

Clearly the force, which is the rate of absorption of momentum from the EM field, is equal to the absorbed energy, given by Eq.(10), divided by the speed $c$ of light in vacuum. Finally, the torque-density exerted on the particle by the EM field is given by

$$\begin{aligned} \boldsymbol{T} &= \boldsymbol{P}\times\boldsymbol{E} \\ &= \varepsilon_0|\chi|E_0^2[-\sin(k_0 z_0 - \omega t + \varphi)\cos(k_0 z_0 - \omega t) + \cos(k_0 z_0 - \omega t + \varphi)\sin(k_0 z_0 - \omega t)]\hat{\boldsymbol{z}} \\ &= -\varepsilon_0|\chi|E_0^2\sin\varphi\,\hat{\boldsymbol{z}}. \end{aligned} \tag{12}$$

The torque is seen to be equal to the rate of absorption of EM energy, as given by Eq.(10), divided by the frequency $\omega$. The angular momentum of the RCP light is thus equal to its energy divided by $\omega$. The minus sign in the final result of Eq.(12) indicates that, in the case of RCP light, the angular momentum is oriented opposite to the propagation direction of the beam. In the language of quantum optics, a photon of energy $\mathcal{E} = \hbar\omega$ propagating along the $z$-axis, has linear momentum $\boldsymbol{p} = (\hbar\omega/c)\hat{\boldsymbol{z}}$ and angular momentum $\boldsymbol{\mathcal{L}} = \pm\hbar\hat{\boldsymbol{z}}$, with either the plus or the minus sign, depending on whether the photon is left- or right-circularly-polarized. Here $\hbar = h/2\pi$ is the reduced Planck constant.

**5. Oscillating current sheet radiating circularly-polarized plane-waves**. The principle of conservation of angular momentum dictates that a source that emits light which carries angular momentum must somehow overcome the deficit by acquiring an equal but opposite amount of angular momentum. In this section we show that an electric-current-carrying sheet that radiates circularly-polarized light will experience a torque that imparts to the sheet an equal but opposite amount of mechanical angular momentum.

With reference to Fig.1, let there exist, within the $xy$-plane at $z=0$, an electric current-density $\boldsymbol{\mathcal{J}}(\boldsymbol{r},t)=\mathcal{J}_{s0}\delta(z)[\sin(\omega t)\,\hat{\boldsymbol{x}}+\cos(\omega t)\,\hat{\boldsymbol{y}}]$. The sheet radiates two circularly polarized plane-waves, one propagating along the positive $z$-axis, the other along the negative $z$-axis, whose electric and magnetic fields are given by

$$\boldsymbol{E}(\boldsymbol{r},t)=\begin{cases}+½Z_0\mathcal{J}_{s0}[\sin(k_0z-\omega t)\,\hat{\boldsymbol{x}}-\cos(k_0z-\omega t)\,\hat{\boldsymbol{y}}]; & z\geq 0,\\ -½Z_0\mathcal{J}_{s0}[\sin(k_0z+\omega t)\,\hat{\boldsymbol{x}}+\cos(k_0z+\omega t)\,\hat{\boldsymbol{y}}]; & z\leq 0.\end{cases} \quad (13a)$$

$$\boldsymbol{H}(\boldsymbol{r},t)=\begin{cases}+½\mathcal{J}_{s0}[\cos(k_0z-\omega t)\,\hat{\boldsymbol{x}}+\sin(k_0z-\omega t)\,\hat{\boldsymbol{y}}]; & z>0,\\ -½\mathcal{J}_{s0}[\cos(k_0z+\omega t)\,\hat{\boldsymbol{x}}-\sin(k_0z+\omega t)\,\hat{\boldsymbol{y}}]; & z<0.\end{cases} \quad (13b)$$

In the above equations, as before, $k_0=\omega/c=2\pi/\lambda_0$ is the wavenumber, while $Z_0$ is the impedance of free space. It is readily verified that the $E$-field is continuous at $z=0$, whereas the $H$-field exhibits a discontinuity $\mathcal{J}_{s0}[\cos(\omega t)\,\hat{\boldsymbol{x}}-\sin(\omega t)\,\hat{\boldsymbol{y}}]$ across the current-carrying sheet. This is consistent with the boundary conditions imposed by Maxwell's equations on the $\boldsymbol{E}$ and $\boldsymbol{H}$ fields at the sheet, namely, the continuity of the tangential $E$-field and the discontinuity of the tangential $H$-field in the presence of a surface current. The rate of flow of EM energy radiated by the sheet is given by the Poynting vector $\boldsymbol{S}(\boldsymbol{r},t)=\boldsymbol{E}(\boldsymbol{r},t)\times\boldsymbol{H}(\boldsymbol{r},t)$, as follows:

$$\boldsymbol{S}(\boldsymbol{r},t)=\begin{cases}+¼Z_0\mathcal{J}_{s0}^2\hat{\boldsymbol{z}}; & z>0,\\ -¼Z_0\mathcal{J}_{s0}^2\hat{\boldsymbol{z}}; & z<0.\end{cases} \quad (14)$$

Note that the total radiated energy per unit cross-sectional area per unit time is precisely equal to $-\iiint\boldsymbol{E}(\boldsymbol{r},t)\cdot\boldsymbol{\mathcal{J}}(\boldsymbol{r},t)\mathrm{d}x\mathrm{d}y\mathrm{d}z$, where the integral is taken over a unit area of the $xy$-plane. The EM momentum of the radiated field is along the positive $z$-axis on the right-hand-side, and along the negative $z$-axis on the left-hand-side of the sheet; therefore, the Lorentz force-density $\boldsymbol{F}(\boldsymbol{r},t)=\rho\boldsymbol{E}+\boldsymbol{\mathcal{J}}\times\boldsymbol{B}$ acting on the sheet must be zero, which is readily verified. (In the formula for the Lorentz force-density, $\rho(\boldsymbol{r},t)$ is the electric charge-density, which, in the present example, is equal to zero.)

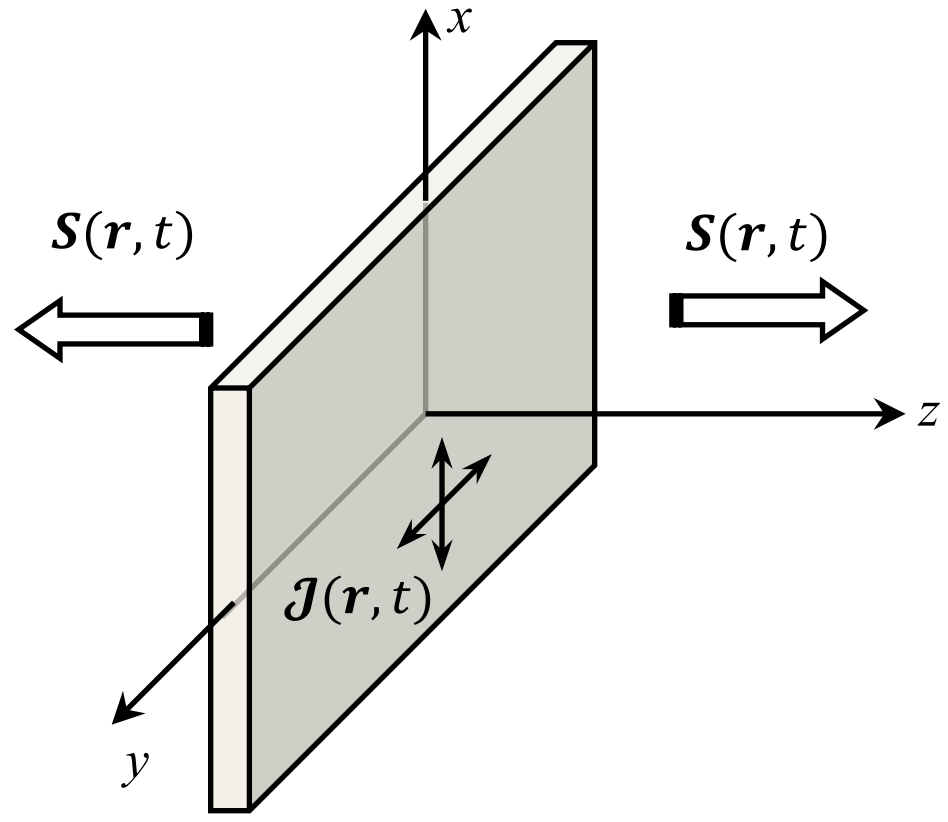

**Fig.1**. The oscillating current-density confined to the $xy$-plane, $\boldsymbol{\mathcal{J}}(\boldsymbol{r},t)=\mathcal{J}_{s0}\delta(z)[\sin(\omega t)\,\hat{\boldsymbol{x}}+\cos(\omega t)\,\hat{\boldsymbol{y}}]$, radiates a right-circularly-polarized plane-wave along the positive $z$-axis, and a left-circularly-polarized plane-wave along the negative $z$-axis. The angular momentum carried away by the EM field is balanced by the mechanical torque experienced by the current-carrying sheet.

Equation (13a) reveals that the plane-wave residing in the $z>0$ half-space is right-circularly polarized, whereas that in the $z<0$ region is left-circularly polarized. This means that, in both regions, the angular momentum of the radiated field is aligned with the negative $z$-axis. Given that each photon carries $\hbar\omega$ of energy and, in the case of circularly-polarized photons, a (spin) angular momentum of $\pm\hbar$, the EM angular momentum per unit-area per unit-time leaving the radiant sheet is

$$\boldsymbol{\mathcal{L}} = -2\boldsymbol{S}/\omega = -(½Z_0\mathcal{J}_{s0}^2/\omega)\hat{\boldsymbol{z}}. \tag{15}$$

We now write the current-density of a finite-area square sheet (length = width = $L$) as follows:

$$\boldsymbol{\mathcal{J}}(\boldsymbol{r},t) = \mathcal{J}_{s0}\text{Rect}(x/L)\text{Rect}(y/L)\delta(z)[\sin(\omega t)\,\hat{\boldsymbol{x}} + \cos(\omega t)\,\hat{\boldsymbol{y}}]. \tag{16}$$

Here $\text{Rect}(x)$ is a rectangular function, which equals 1.0 when $|x| < ½$, and zero otherwise. The charge-current continuity equation $\boldsymbol{\nabla}\cdot\boldsymbol{\mathcal{J}} + (\partial\rho/\partial t) = 0$ yields the electrical charge-density produced at the edges of the sheet as

$$\rho(\boldsymbol{r},t) = (\mathcal{J}_{s0}/\omega)[\delta(x+½L) - \delta(x-½L)]\text{Rect}(y/L)\delta(z)\cos(\omega t)$$
$$-(\mathcal{J}_{s0}/\omega)\text{Rect}(x/L)[\delta(y+½L) - \delta(y-½L)]\delta(z)\sin(\omega t). \tag{17}$$

The $E$-field of Eq.(13a) acting on the charge-density $\rho(\boldsymbol{r},t)$ residing at the edges of the sheet gives rise to a Lorentz force-density $\rho(\boldsymbol{r},t)\boldsymbol{E}(\boldsymbol{r},t)$, which then produces a torque acting on the sheet, namely,

$$\boldsymbol{T}(t) = ½(Z_0\mathcal{J}_{s0}^2/\omega)L^2[\sin^2(\omega t) + \cos^2(\omega t)]\hat{\boldsymbol{z}}. \tag{18}$$

It is seen that the torque acting per unit-area of the sheet is precisely equal in magnitude and opposite in direction to the EM angular momentum per unit-area per unit-time that is radiated away, confirming thereby the conservation of angular momentum.

**6. Vector cylindrical harmonics**. The eigen-mode solutions of Maxwell's equations in linear, homogeneous, isotropic, and cylindrically-symmetric systems are known as vector cylindrical harmonics [28]. These solutions have been extensively studied in the context of cylindrical waveguides, in general [29], and optical fibers [30,31] as well as metallic nanowires [32], in particular. The propagation vector (or $k$-vector) associated with a cylindrical harmonic typically has a radial component $k_r$ and an axial component $k_z$. When $|k_r| \ll |k_z|$ the eigen-mode is said to be in the paraxial regime. We will examine the angular momentum of vector cylindrical harmonics with arbitrary $k_r \neq 0$ and $k_z \neq 0$ in Sec.7. In the present section, however, we limit our discussion to the case of $k_z = 0$, which is sometimes referred to as the anti-paraxial regime. This restriction to $k_z = 0$, however, is solely for mathematical convenience and algebraic simplicity, as the more general case of $k_z \neq 0$ can be treated along the same lines as those pursued below, albeit the treatment would involve somewhat tedious calculations.

Consider an infinitely-long, thin, hollow right-circular cylinder of radius $R_0$ carrying the current-density $\boldsymbol{\mathcal{J}}(r,\phi,z) = \mathcal{J}_{s0}\delta(r-R_0)\exp[\mathrm{i}(m\phi-\omega t)]\,\hat{\boldsymbol{z}}$ on its surface. Here $\mathcal{J}_{s0}$ is the amplitude of the surface-current-density, $\delta(\cdot)$ is Dirac's delta-function, $\omega$ is the oscillation frequency, and $m = 0, \pm1, \pm2, \cdots$ is an arbitrary integer. The EM fields inside and outside the cylinder, constituting a transverse magnetic (TM) mode of order $m$, are found to be

$$\boldsymbol{E}(\boldsymbol{r},t) = \begin{bmatrix} E_0 J_m(k_0 r)\hat{\boldsymbol{z}} \\ \\ E_1\mathcal{H}_m^{(1)}(k_0 r)\hat{\boldsymbol{z}} \end{bmatrix} \times \exp[\mathrm{i}(m\phi-\omega t)]\,; \quad \begin{matrix} (r < R_0), \\ \\ (r > R_0). \end{matrix} \tag{19}$$

$$\boldsymbol{H}(\boldsymbol{r},t)=\begin{bmatrix}(mE_0/Z_0k_0r)\,J_m(k_0r)\hat{\boldsymbol{r}}+\mathrm{i}(E_0/Z_0)\,\dot{J}_m(k_0r)\hat{\boldsymbol{\phi}} & (r<R_0),\\ (mE_1/Z_0k_0r)\,\mathcal{H}_m^{(1)}(k_0r)\hat{\boldsymbol{r}}+\mathrm{i}(E_1/Z_0)\,\dot{\mathcal{H}}_m^{(1)}(k_0r)\hat{\boldsymbol{\phi}} & (r>R_0).\end{bmatrix}\times\exp[\mathrm{i}(m\phi-\omega t)]\,; \tag{20}$$

In the above equations, $E_0$ and $E_1$ are the $E$-field amplitudes inside and outside the cylinder, $k_0=\omega/c$ is the wavenumber, $c$ is the speed of light in vacuum, $Z_0$ is the impedance of free space, $J_m(\rho)$ is a Bessel function of the 1st kind, order $m$, $\mathcal{H}_m^{(1)}(\rho)=J_m(\rho)+\mathrm{i}Y_m(\rho)$ is a Bessel function of the 3rd kind, type 1, order $m$ (also known as a type 1 Hankel function), and the over-dots of $\dot{J}_m(\rho)$ and $\dot{\mathcal{H}}_m^{(1)}(\rho)$ indicate differentiation with respect to $\rho$.

At the cylinder surface, the boundary conditions for $E_\parallel$ and $H_\parallel$ (and also for $B_\perp$, where $\boldsymbol{B}=\mu_0\boldsymbol{H}$) yield

$$E_1\mathcal{H}_m^{(1)}(k_0R_0)=E_0J_m(k_0R_0), \tag{21a}$$

$$\mathrm{i}(E_1/Z_0)\dot{\mathcal{H}}_m^{(1)}(k_0R_0)-\mathrm{i}(E_0/Z_0)\dot{J}_m(k_0R_0)=\mathcal{J}_{s0}. \tag{21b}$$

The above equations are readily solved for the $E$-field amplitudes $E_0$ and $E_1$, as follows:

$$E_0=-\tfrac{1}{2}\pi k_0R_0Z_0\mathcal{J}_{s0}\mathcal{H}_m^{(1)}(k_0R_0), \tag{22a}$$

$$E_1=-\tfrac{1}{2}\pi k_0R_0Z_0\mathcal{J}_{s0}\,J_m(k_0R_0). \tag{22b}$$

The following Bessel function identities have been used in the above derivation [33,34]:

$$Y_m(\rho)J_{m+1}(\rho)-J_m(\rho)Y_{m+1}(\rho)=2/(\pi\rho). \tag{23}$$

$$\dot{\mathcal{Z}}_m(\rho)=(m/\rho)\mathcal{Z}_m(\rho)-\mathcal{Z}_{m+1}(\rho). \tag{24}$$

In the latter identity, $\mathcal{Z}_m$ could represent a Bessel function of the first kind, $J_m$, a Bessel function of the second kind, $Y_m$, or a Bessel function of the third kind, either type 1 or type 2, $\mathcal{H}_m^{(1,2)}=J_m\pm\mathrm{i}Y_m$.

**Example 1**. In the case of a hollow, infinitely long, perfectly-electrically-conducting cylinder shown in Fig.2, the current-density $\boldsymbol{\mathcal{J}}(\boldsymbol{r},t)$ will be confined to the inner cylindrical surface, and both $\boldsymbol{E}$ and $\boldsymbol{H}$ fields within the metallic shell as well as outside the cylinder vanish. Therefore, $E_1=0$, which implies that $k_0R_0$ must be a zero of the Bessel function $J_m(\rho)$. This constraint ensures that Maxwell's boundary conditions at the inner surface of the cylinder are satisfied, namely, $\boldsymbol{E}_\parallel$ and $\boldsymbol{B}_\perp$ vanish, while $\boldsymbol{H}_\parallel=H_\phi\hat{\boldsymbol{\phi}}=-\mathcal{J}_{s0}\hat{\boldsymbol{\phi}}$, as shown below.

$$\begin{aligned}H_\phi(r=R_0,\phi,z,t)&=\mathrm{i}(E_0/Z_0)\,\dot{J}_m(k_0R_0)\exp[\mathrm{i}(m\phi-\omega t)]\\&=-\tfrac{1}{2}\mathrm{i}\pi k_0R_0\,\mathcal{J}_{s0}\mathcal{H}_m^{(1)}(k_0R_0)\,\dot{J}_m(k_0R_0)\exp[\mathrm{i}(m\phi-\omega t)]\\&=\tfrac{1}{2}\pi k_0R_0\,\mathcal{J}_{s0}Y_m(k_0R_0)\,\dot{J}_m(k_0R_0)\exp[\mathrm{i}(m\phi-\omega t)]\\&=-\tfrac{1}{2}\pi k_0R_0\,\mathcal{J}_{s0}Y_m(k_0R_0)\,J_{m+1}(k_0R_0)\exp[\mathrm{i}(m\phi-\omega t)]\\&=-\mathcal{J}_{s0}\exp[\mathrm{i}(m\phi-\omega t)].\end{aligned} \tag{25}$$

**Fig.2**. An infinitely long, hollow, perfectly-electrically-conducting, right-circular cylinder carries the current-density $\boldsymbol{\mathcal{J}}(r,\phi,z) = \mathcal{J}_{s0}\delta(r-R_0)\exp[\mathrm{i}(m\phi-\omega t)]\,\hat{\boldsymbol{z}}$ on its interior surface. Considering that the EM field in the metallic shell must vanish, the product $k_0R_0$ of the wavenumber $k_0 = 2\pi/\lambda_0$ and the radius $R_0$ of the cylindrical cavity must be a zero of the Bessel function $J_m(\rho)$. Within a unit-length of the cavity, the trapped EM field has orbital angular momentum $\boldsymbol{\mathcal{L}}$ given by Eq.(26), and energy $\mathcal{E}$ given by Eq.(27). The ratio $\boldsymbol{\mathcal{L}}/\mathcal{E}$ is thus equal to $(m/\omega)\hat{\boldsymbol{z}}$, independently of $R_0$.

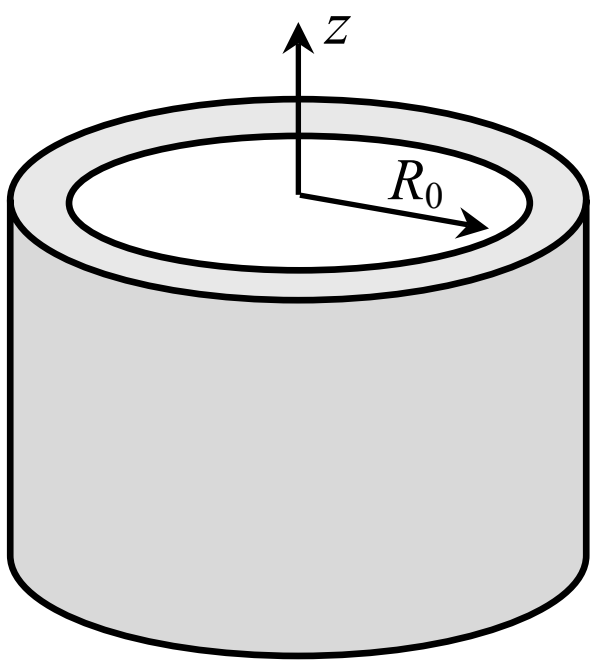

The phase-factor $\exp(\mathrm{i}m\phi)$ of the trapped EM field inside the cylindrical cavity indicates that the field has vorticity of order $m$. The angular momentum of the trapped vortex per unit-length of the cylinder is directly computed from a knowledge of the Poynting vector $\boldsymbol{S}(\boldsymbol{r},t) = \mathrm{Re}[\boldsymbol{E}(\boldsymbol{r},t)] \times \mathrm{Re}[\boldsymbol{H}(\boldsymbol{r},t)]$, as follows:

$$\boldsymbol{\mathcal{L}}(t) = \int_{r=0}^{R_0}\int_{\phi=0}^{2\pi}\int_{z=0}^{1} \boldsymbol{r} \times [\boldsymbol{S}(\boldsymbol{r},t)/c^2] r\mathrm{d}r\mathrm{d}\phi\mathrm{d}z$$

$$= [\tfrac{1}{2}\pi k_0 R_0 Z_0 \mathcal{J}_{s0} Y_m(k_0R_0)/c]^2 \int_{r=0}^{R_0}\int_{\phi=0}^{2\pi} r\hat{\boldsymbol{r}} \times \{J_m(k_0r)\sin(m\phi-\omega t)\,\hat{\boldsymbol{z}}$$

$$\times [(m/Z_0k_0r)\,J_m(k_0r)\sin(m\phi-\omega t)\,\hat{\boldsymbol{r}} + (1/Z_0)\dot{J}_m(k_0r)\cos(m\phi-\omega t)\,\hat{\boldsymbol{\phi}}]\}\, r\mathrm{d}r\mathrm{d}\phi$$

$$= [m\mu_0\pi^2k_0^2R_0^2\mathcal{J}_{s0}^2Y_m^2(k_0R_0)\hat{\boldsymbol{z}}/(4\omega)] \int_{r=0}^{R_0}\int_{\varphi=0}^{2\pi} rJ_m^2(k_0r)\sin^2(m\varphi-\omega t)\,\mathrm{d}r\mathrm{d}\varphi$$

$$= [m\mu_0\pi^3k_0^2R_0^2\mathcal{J}_{s0}^2Y_m^2(k_0R_0)\hat{\boldsymbol{z}}/(4\omega)] \int_0^{R_0} rJ_m^2(k_0r)\mathrm{d}r$$

$$= [m\mu_0\pi^3k_0^2R_0^2\mathcal{J}_{s0}^2Y_m^2(k_0R_0)\hat{\boldsymbol{z}}/(4\omega k_0^2)] \int_0^{k_0R_0} \rho J_m^2(\rho)\mathrm{d}\rho \qquad \boxed{m=0 \;\rightarrow\; J_{-1}(\cdot) = -J_1(\cdot).}$$

$$= [m\mu_0\pi^3k_0^2R_0^2\mathcal{J}_{s0}^2Y_m^2(k_0R_0)\hat{\boldsymbol{z}}/(8\omega k_0^2)][\rho^2J_m^2(\rho) - \rho^2J_{m-1}(\rho)J_{m+1}(\rho)]_0^{k_0R_0}$$

$$= [m\mu_0\pi^3k_0^2R_0^4\mathcal{J}_{s0}^2\hat{\boldsymbol{z}}/(8\omega)][J_m^2(k_0R_0) - J_{m-1}(k_0R_0)J_{m+1}(k_0R_0)]Y_m^2(k_0R_0) \;\leftarrow\; \boxed{J_m(k_0R_0) = 0}$$

$$= m\pi R_0^2\mu_0\mathcal{J}_{s0}^2\hat{\boldsymbol{z}}/(2\omega). \qquad (26)$$

In addition to Eq.(23), the identity $\int \rho J_m^2(\rho)\mathrm{d}\rho = \tfrac{1}{2}\rho^2[J_m^2(\rho) - J_{m-1}(\rho)J_{m+1}(\rho)]$ has been used in the above derivation [34]. Next, we compute the EM energy content of the trapped mode per unit-length of the cylinder, as follows:

$$\mathcal{E}(t) = \int_{r=0}^{R_0}\int_{\phi=0}^{2\pi}\int_{z=0}^{1}\{\tfrac{1}{2}\varepsilon_0\mathrm{Re}^2[\boldsymbol{E}(\boldsymbol{r},t)] + \tfrac{1}{2}\mu_0\mathrm{Re}^2[\boldsymbol{H}(\boldsymbol{r},t)]\}\, r\mathrm{d}r\mathrm{d}\phi\mathrm{d}z$$

$$= \tfrac{1}{8}\mu_0\pi^2k_0^2R_0^2\mathcal{J}_{s0}^2Y_m^2(k_0R_0)\int_{r=0}^{R_0}\int_{\phi=0}^{2\pi}\{[J_m^2(k_0r) + (m/k_0r)^2J_m^2(k_0r)]\sin^2(m\phi-\omega t)$$

$$+\dot{J}_m^2(k_0r)\cos^2(m\phi-\omega t)\}\, r\mathrm{d}r\mathrm{d}\phi$$

$$= \tfrac{1}{8}\mu_0\pi^3k_0^2R_0^2\mathcal{J}_{s0}^2Y_m^2(k_0R_0)\int_0^{R_0}[J_m^2(k_0r) + (m/k_0r)^2J_m^2(k_0r) + \dot{J}_m^2(k_0r)]r\mathrm{d}r$$

$$= \tfrac{1}{8}\mu_0\pi^3k_0^2R_0^2\mathcal{J}_{s0}^2Y_m^2(k_0R_0)$$

$$\times \int_0^{R_0}\{J_m^2(k_0r) + [(m/k_0r)J_m(k_0r) - \dot{J}_m(k_0r)]^2 + (2m/k_0r)J_m(k_0r)\dot{J}_m(k_0r)\}r\mathrm{d}r$$

$$= \tfrac{1}{8}\mu_0\pi^3 k_0^2 R_0^2 \mathcal{J}_{s0}^2 Y_m^2(k_0R_0)\int_0^{R_0}\{rJ_m^2(k_0r) + rJ_{m+1}^2(k_0r) + (2m/k_0)J_m(k_0r)\dot{J}_m(k_0r)\}\mathrm{d}r$$

$$= \tfrac{1}{8}\mu_0\pi^3 k_0^2 R_0^2 \mathcal{J}_{s0}^2 Y_m^2(k_0R_0)\{\tfrac{1}{2}R_0^2[J_m^2(k_0R_0) - J_{m-1}(k_0R_0)J_{m+1}(k_0R_0)] \quad \leftarrow \boxed{J_m(k_0R_0) = 0}$$

$$+\tfrac{1}{2}R_0^2[J_{m+1}^2(k_0R_0) - J_m(k_0R_0)J_{m+2}(k_0R_0)] + (m/k_0^2)J_m^2(k_0R_0)\}$$

$$= \tfrac{1}{16}\mu_0\pi^3 k_0^2 R_0^4 \mathcal{J}_{s0}^2[J_{m+1}^2(k_0R_0) - J_{m-1}(k_0R_0)J_{m+1}(k_0R_0)]Y_m^2(k_0R_0) = \tfrac{1}{2}\pi R_0^2\mu_0\mathcal{J}_{s0}^2. \tag{27}$$

A comparison of Eqs.(26) and (27) reveals that $\boldsymbol{\mathcal{L}}(t)/\mathcal{E}(t) = (m/\omega)\hat{\boldsymbol{z}}$, which is consistent with the quantum-optical picture of photons of vorticity $m$ trapped in the cavity, each having energy $\hbar\omega$ and (orbital) angular momentum $m\hbar\hat{\boldsymbol{z}}$. Also of interest is the average radiation pressure on the interior surface of the cylinder, which is readily computed as follows:

$$\mathcal{P}(t) = (2\pi R_0)^{-1}\int_{r=R_0^-}^{R_0^+}\int_{\phi=0}^{2\pi}|\boldsymbol{\mathcal{J}}(r,\phi,z,t)\times\mu_0\boldsymbol{H}(r,\phi,z,t)|r\mathrm{d}r\mathrm{d}\phi$$

$$= (\mu_0\mathcal{J}_{s0}^2/4\pi)\int_{\varphi=0}^{2\pi}\cos^2(m\phi-\omega t)\,\mathrm{d}\phi = \begin{cases}\tfrac{1}{2}\mu_0\mathcal{J}_{s0}^2\cos^2(\omega t); & m = 0,\\ \tfrac{1}{4}\mu_0\mathcal{J}_{s0}^2; & m \neq 0.\end{cases} \tag{28}$$

Comparing Eqs.(27) and (28), we find that $\langle\mathcal{P}\rangle = \mathcal{E}/(2\pi R_0^2)$, which is independent of the order $m$ of vorticity (and, therefore, of the angular momentum) of the trapped beam. The circulation of the EM energy around the $z$-axis is thus seen to have no effect on the pressure exerted on the walls of the cavity by the trapped field.

**Example 2**. Consider the pair of infinitely long, perfectly-electrically-conducting, concentric cylinders shown in Fig.3. The solid inner cylinder has radius $R_0$ and carries an electrical current-density $\mathcal{J}_{s0}\delta(r-R_0)\exp[\mathrm{i}(m\phi-\omega t)]\,\hat{\boldsymbol{z}}$ on its exterior surface. The hollow outer cylinder whose inside radius is $R_1$ carries a current-density $\mathcal{J}_{s1}\delta(r-R_1)\exp[\mathrm{i}(m\phi-\omega t)]\,\hat{\boldsymbol{z}}$ on its interior surface. The EM field will now be confined to the empty gap between the two cylinders, i.e., $R_0 < r < R_1$, provided that

$$R_0\mathcal{J}_{s0}\,J_m(k_0R_0) = -R_1\mathcal{J}_{s1}\,J_m(k_0R_1), \tag{29a}$$

$$R_0\mathcal{J}_{s0}Y_m(k_0R_0) = -R_1\mathcal{J}_{s1}Y_m(k_0R_1). \tag{29b}$$

The above equations will be satisfied when

$$\frac{J_m(k_0R_0)}{J_m(k_0R_1)} = \frac{Y_m(k_0R_0)}{Y_m(k_0R_1)} = -\frac{R_1\mathcal{J}_{s1}}{R_0\mathcal{J}_{s0}}. \tag{30}$$

The first of the above equations places a constraint on $R_0$ and $R_1$, since the gap between the cylinders cannot be arbitrary—it must accommodate a certain number of wavelengths. The second part of Eq.(30) then specifies the relative magnitude of the two current-densities. Subsequently, we may write the EM fields in the region between the two cylinders as follows:

$$E_z(\boldsymbol{r},t) = -\tfrac{1}{2}\pi Z_0 k_0\left[R_0\mathcal{J}_{s0}\,J_m(k_0R_0)\mathcal{H}_m^{(1)}(k_0r) + R_1\mathcal{J}_{s1}\mathcal{H}_m^{(1)}(k_0R_1)J_m(k_0r)\right]\exp[\mathrm{i}(m\phi-\omega t)]$$

$$= -\tfrac{1}{2}\pi Z_0\mathcal{J}_{s0}k_0R_0\left[J_m(k_0R_0)\mathcal{H}_m^{(1)}(k_0r) - \mathcal{H}_m^{(1)}(k_0R_0)J_m(k_0r)\right]\exp[\mathrm{i}(m\phi-\omega t)]$$

$$= -\tfrac{1}{4}\pi Z_0 \mathcal{J}_{s0} k_0 R_0 \left[\mathcal{H}_m^{(2)}(k_0R_0)\mathcal{H}_m^{(1)}(k_0r) - \mathcal{H}_m^{(1)}(k_0R_0)\mathcal{H}_m^{(2)}(k_0r)\right] \exp[\mathrm{i}(m\phi-\omega t)]. \qquad (31)$$

$$H_r(\boldsymbol{r},t) = -\tfrac{1}{2}m\pi r^{-1}\left[R_0\mathcal{J}_{s0}J_m(k_0R_0)\mathcal{H}_m^{(1)}(k_0r) + R_1\mathcal{J}_{s1}\mathcal{H}_m^{(1)}(k_0R_1)J_m(k_0r)\right]\exp[\mathrm{i}(m\phi-\omega t)]$$

$$= -\tfrac{1}{4}m\pi(\mathcal{J}_{s0}R_0/r)\left[\mathcal{H}_m^{(2)}(k_0R_0)\mathcal{H}_m^{(1)}(k_0r) - \mathcal{H}_m^{(1)}(k_0R_0)\mathcal{H}_m^{(2)}(k_0r)\right]\exp[\mathrm{i}(m\phi-\omega t)]. \qquad (32)$$

$$H_\phi(\boldsymbol{r},t) = -\tfrac{1}{2}\mathrm{i}\pi k_0\left[R_0\mathcal{J}_{s0}J_m(k_0R_0)\dot{\mathcal{H}}_m^{(1)}(k_0r) + R_1\mathcal{J}_{s1}\mathcal{H}_m^{(1)}(k_0R_1)\dot{J}_m(k_0r)\right]\exp[\mathrm{i}(m\phi-\omega t)]$$

$$= -\tfrac{1}{4}\mathrm{i}\pi\mathcal{J}_{s0}k_0R_0\left[\mathcal{H}_m^{(2)}(k_0R_0)\dot{\mathcal{H}}_m^{(1)}(k_0r) - \mathcal{H}_m^{(1)}(k_0R_0)\dot{\mathcal{H}}_m^{(2)}(k_0r)\right]\exp[\mathrm{i}(m\phi-\omega t)]. \qquad (33)$$

The EM field trapped between the two cylinders is now seen to be the superposition of an outward propagating wave, $\mathcal{H}_m^{(1)}(k_0r)$, and an inward propagating wave $\mathcal{H}_m^{(2)}(k_0r)$. Both the $E$-field $E_z$ and the $B$-field component $B_r = \mu_0 H_r$ vanish on the surfaces of the two cylinders, as demanded by Maxwell's boundary conditions. The tangential $H$-field component $H_\phi$, however, is non-zero on the two surfaces, balancing the surface-current-density $\mathcal{J}_{s0}$ of the inner cylinder at $r = R_0$, and $\mathcal{J}_{s1}$ of the outer cylinder at $r = R_1$.

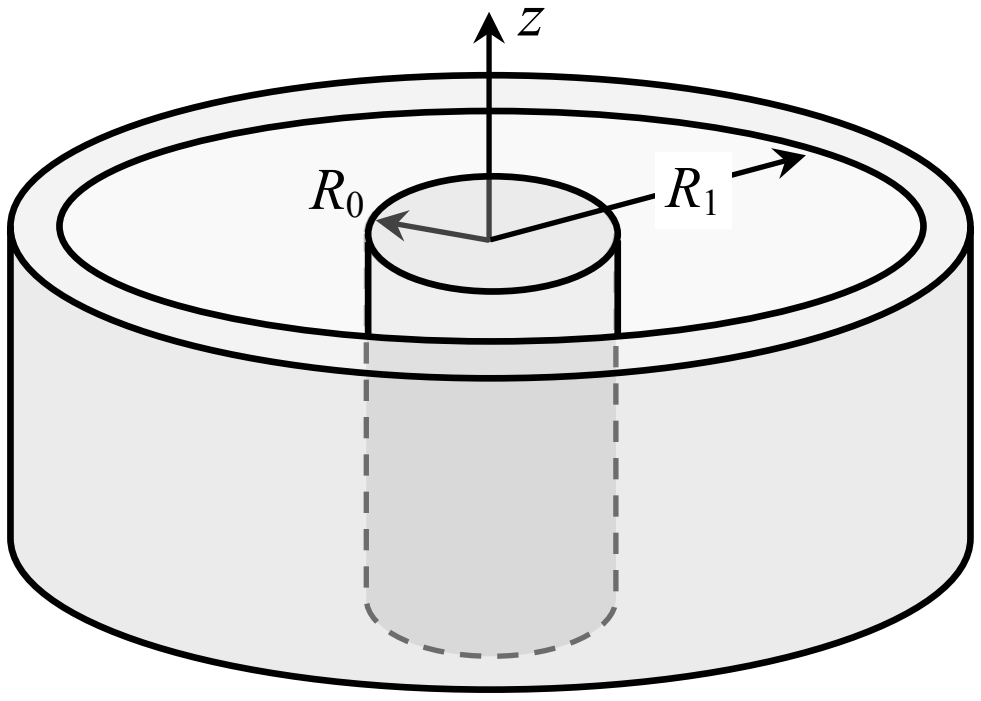

**Fig.3**. A cylindrical harmonic EM field is trapped in the gap between two infinitely long, perfectly-electrically-conducting cylinders. The small cylinder is solid and has radius $R_0$, while the large cylinder is hollow and has inner radius $R_1$. No EM fields reside within the conductors or outside the large cylinder. The current-density $\mathcal{J}_{s0}\delta(r-R_0)\exp[\mathrm{i}(m\phi-\omega t)]\,\hat{\boldsymbol{z}}$ flows on the exterior surface of the small cylinder, whereas the current-density $\mathcal{J}_{s1}\delta(r-R_1)\exp[\mathrm{i}(m\phi-\omega t)]\,\hat{\boldsymbol{z}}$ flows on the interior surface of the large cylinder. The existence of a trapped mode is guaranteed by Eq.(30), which imposes restrictions on the cylinder radii $R_0$ and $R_1$, the wavenumber $k_0$, and the surface-current-densities $\mathcal{J}_{s0}$ and $\mathcal{J}_{s1}$.

The properties of the trapped vortex field between the two cylinders of Fig.3 may now be calculated from Eqs.(30)-(33), along the same lines as those in the preceding example, where the energy and angular momentum of the field inside the hollow cylinder of Fig.2 were computed. The algebra may be more tedious, but the final results will be simple and elegant. For example, the EM angular momentum per unit length trapped between the two cylinders is found to be $\boldsymbol{\mathcal{L}}(t) = m\pi\mu_0(R_1^2\mathcal{J}_{s1}^2 - R_0^2\mathcal{J}_{s0}^2)\hat{\boldsymbol{z}}/(2\omega)$. As a simple special case, consider the case of $R_0 \gg \lambda_0$, for which the following asymptotic formulas in the limit of $\rho \gg 1$ become applicable [34]:

$$J_m(\rho) \cong \sqrt{2/(\pi\rho)}\cos(\rho - \tfrac{1}{2}m\pi - \tfrac{1}{4}\pi), \qquad (34a)$$

$$Y_m(\rho) \cong \sqrt{2/(\pi\rho)}\sin(\rho - \tfrac{1}{2}m\pi - \tfrac{1}{4}\pi). \qquad (34b)$$

The left-hand side of Eq.(30) now yields $\tan(k_0R_0 - \tfrac{1}{2}m\pi - \tfrac{1}{4}\pi) = \tan(k_0R_1 - \tfrac{1}{2}m\pi - \tfrac{1}{4}\pi)$, which requires that $k_0(R_1 - R_0)$ be an integer multiple of $\pi$. In other words, the gap between the two cylinders must be an integer-multiple of $\lambda_0/2$. Under these circumstances, the right-hand-side of Eq.(30) yields $\sqrt{R_0}\mathcal{J}_{s0} = \pm\sqrt{R_1}\mathcal{J}_{s1}$. Considering that the radiation pressure on the inner and outer cylindrical surfaces is given by $\tfrac{1}{4}\mu_0\mathcal{J}_{s0}^2$ and $\tfrac{1}{4}\mu_0\mathcal{J}_{s1}^2$, respectively, it is readily seen that the integrated radiation pressure around the circumference of the inner cylinder is equal to that on the outer cylinder. Moreover, the radiation pressure is seen to be independent of the azimuthal mode number $m$, indicating that the orbital angular momentum of the field has no bearing on the radiation pressure exerted on the inner and outer cylinder surfaces.

**7. Orbital angular momentum of a vector cylindrical harmonic EM wave**. The subject of the present section is the behavior of a small spherical particle exposed to a transverse electric (TE) mode cylindrical harmonic EM wave in free space. In general, the $k$-vector of a cylindrical harmonic can have components $k_r = k_0\sigma_r$ and $k_z = k_0\sigma_z$ along both $\hat{\boldsymbol{r}}$ and $\hat{\boldsymbol{z}}$ directions. Although in the preceding section we limited our discussion to transverse magnetic (TM) modes with $k_z = 0$, here, in order to emphasize the generality of the results, we analyze (from an altogether different perspective) the angular momentum of a TE mode cylindrical harmonic with $k_z \neq 0$. The $\boldsymbol{E}$ and $\boldsymbol{H}$ fields of a TE eigen-mode of integer order $m$ are given by [28]:

$$\boldsymbol{E}(\boldsymbol{r},t) = (mZ_0H_0/k_0r\sigma_r^2)J_m(k_0\sigma_r r)\cos(k_0\sigma_z z + m\phi - \omega t)\,\hat{\boldsymbol{r}}$$
$$-(Z_0H_0/\sigma_r)\dot{J}_m(k_0\sigma_r r)\sin(k_0\sigma_z z + m\phi - \omega t)\,\hat{\boldsymbol{\phi}}. \quad (35)$$

$$\boldsymbol{H}(\boldsymbol{r},t) = (\sigma_z H_0/\sigma_r)\dot{J}_m(k_0\sigma_r r)\sin(k_0\sigma_z z + m\phi - \omega t)\,\hat{\boldsymbol{r}}$$
$$+(m\sigma_z H_0/k_0 r\sigma_r^2)J_m(k_0\sigma_r r)\cos(k_0\sigma_z z + m\phi - \omega t)\,\hat{\boldsymbol{\phi}}$$
$$-H_0 J_m(k_0\sigma_r r)\cos(k_0\sigma_z z + m\phi - \omega t)\,\hat{\boldsymbol{z}}. \quad (36)$$

Here $\boldsymbol{r} = (r,\phi,z)$ is the position vector in cylindrical coordinates, $J_m(\rho)$ is a Bessel function of the first kind, order $m$, and $\dot{J}_m(\rho)$ is the derivative of $J_m(\rho)$ with respect to $\rho$. The only component of the field along the $z$-axis is $H_z$, whose amplitude is specified as $H_0$. The oscillation frequency is $\omega$, the wavenumber is $k_0 = \omega/c$, and the wave-vector is written as $\boldsymbol{k} = k_0(\sigma_r\hat{\boldsymbol{r}} + \sigma_z\hat{\boldsymbol{z}})$, where $\sigma_r$ and $\sigma_z$ are both real and positive, with $\sigma_r^2 + \sigma_z^2 = 1$. As usual, $Z_0$ is the impedance of free space, and the integer $m$ denotes the azimuthal order of the mode.

Let a small polarizable particle having susceptibility $\varepsilon_0|\chi|\exp(\mathrm{i}\varphi)$ be placed at a fixed point $\boldsymbol{r} = (r,\phi,z)$ in space. The induced polarization (i.e., electric dipole moment density) of the particle will be

$$\boldsymbol{P}(\boldsymbol{r},t) = \varepsilon_0|\chi|(mZ_0H_0/k_0r\sigma_r^2)J_m(k_0\sigma_r r)\cos(k_0\sigma_z z + m\phi - \omega t + \varphi)\,\hat{\boldsymbol{r}}$$
$$-\varepsilon_0|\chi|(Z_0H_0/\sigma_r)\dot{J}_m(k_0\sigma_r r)\sin(k_0\sigma_z z + m\phi - \omega t + \varphi)\,\hat{\boldsymbol{\phi}}. \quad (37)$$

The particle absorbs EM energy from the field at a (time-averaged) rate that is given by

$$\frac{\mathrm{d}\mathcal{E}}{\mathrm{d}t} = \boldsymbol{E}\cdot\frac{\partial \boldsymbol{P}}{\partial t} = \varepsilon_0|\chi|\omega Z_0^2H_0^2(m/k_0r\sigma_r^2)^2J_m^2(k_0\sigma_r r)\cos(k_0\sigma_z z + m\phi - \omega t)\sin(k_0\sigma_z z + m\phi - \omega t + \varphi)$$
$$-\varepsilon_0|\chi|\omega Z_0^2H_0^2(1/\sigma_r)^2\dot{J}_m^2(k_0\sigma_r r)\sin(k_0\sigma_z z + m\phi - \omega t)\cos(k_0\sigma_z z + m\phi - \omega t + \varphi).$$

$$\langle \mathrm{d}\mathcal{E}/\mathrm{d}t\rangle = \langle \boldsymbol{E}\cdot\partial\boldsymbol{P}/\partial t\rangle = (|\chi|\mu_0H_0^2\omega/2\sigma_r^2)[(m/k_0r\sigma_r)^2J_m^2(k_0\sigma_r r) + \dot{J}_m^2(k_0\sigma_r r)]\sin\varphi. \quad (38)$$

The EM field exerts on the particle the Lorentz force-density $\boldsymbol{F} = (\boldsymbol{P}\cdot\boldsymbol{\nabla})\boldsymbol{E} + (\partial\boldsymbol{P}/\partial t)\times\mu_0\boldsymbol{H}$ and the Lorentz torque-density $\boldsymbol{T} = \boldsymbol{r}\times\boldsymbol{F} + \boldsymbol{P}\times\boldsymbol{E}$. The various terms appearing in these expressions can be computed straightforwardly as follows:

$$(\boldsymbol{P}\cdot\boldsymbol{\nabla})\boldsymbol{E}|_r = \mu_0|\chi|(m^2H_0^2/k_0r^2\sigma_r^3)J_m(k_0\sigma_r r)[\dot{J}_m(k_0\sigma_r r) - (1/k_0r\sigma_r)J_m(k_0\sigma_r r)]$$
$$\times\cos(k_0\sigma_z z + m\phi - \omega t + \varphi)\cos(k_0\sigma_z z + m\phi - \omega t)$$
$$+\mu_0|\chi|(m^2H_0^2/k_0r^2\sigma_r^3)J_m(k_0\sigma_r r)\dot{J}_m(k_0\sigma_r r)\sin(k_0\sigma_z z + m\phi - \omega t + \varphi)\sin(k_0\sigma_z z + m\phi - \omega t)$$
$$-\mu_0|\chi|(H_0^2/r\sigma_r^2)\dot{J}_m^2(k_0\sigma_r r)\sin(k_0\sigma_z z + m\phi - \omega t + \varphi)\sin(k_0\sigma_z z + m\phi - \omega t).$$

$$(\boldsymbol{P}\cdot\boldsymbol{\nabla})\boldsymbol{E}|_\phi = -\mu_0|\chi|H_0^2(m/r\sigma_r^2)\{J_m(k_0\sigma_r r)\ddot{J}_m(k_0\sigma_r r)\cos(k_0\sigma_z z+m\phi-\omega t+\varphi)\sin(k_0\sigma_z z+m\phi-\omega t)$$
$$+(1/k_0\sigma_r r)J_m(k_0\sigma_r r)\dot{J}_m(k_0\sigma_r r)\sin(k_0\sigma_z z+m\phi-\omega t+\varphi)\cos(k_0\sigma_z z+m\phi-\omega t)$$
$$-\dot{J}_m^2(k_0\sigma_r r)\sin(k_0\sigma_z z+m\phi-\omega t+\varphi)\cos(k_0\sigma_z z+m\phi-\omega t)\}.$$

$$\langle\boldsymbol{r}\times(\boldsymbol{P}\cdot\boldsymbol{\nabla})\boldsymbol{E}\rangle|_z = ½\mu_0|\chi|H_0^2(m/\sigma_r^2)\big[J_m(k_0\sigma_r r)\ddot{J}_m(k_0\sigma_r r)-(1/k_0\sigma_r r)J_m(k_0\sigma_r r)\dot{J}_m(k_0\sigma_r r)+\dot{J}_m^2(k_0\sigma_r r)\big]$$
$$\times\sin\varphi$$
$$= ½\mu_0|\chi|H_0^2(m/\sigma_r^2)\big\{J_m(k_0\sigma_r r)\{[(m/k_0\sigma_r r)^2-1]J_m(k_0\sigma_r r)-(1/k_0\sigma_r r)\dot{J}_m(k_0\sigma_r r)\}$$
$$-\big[(1/k_0 r\sigma_r)J_m(k_0\sigma_r r)-\dot{J}_m(k_0\sigma_r r)\big]\dot{J}_m(k_0\sigma_r r)\big\}\sin\varphi.$$

$$\langle\boldsymbol{r}\times(\boldsymbol{P}\cdot\boldsymbol{\nabla})\boldsymbol{E}\rangle|_z = ½(m|\chi|\mu_0 H_0^2/\sigma_r^2)\big\{[(m/k_0\sigma_r r)^2-1]J_m^2(k_0\sigma_r r)+\dot{J}_m^2(k_0\sigma_r r)$$
$$-(2/k_0\sigma_r r)J_m(k_0\sigma_r r)\dot{J}_m(k_0\sigma_r r)\big\}\sin\varphi. \quad (39)$$

$$\frac{\partial\boldsymbol{P}}{\partial t}\times\mu_0\boldsymbol{H} = \mu_0|\chi|H_0^2(\omega/c\sigma_r)\{(m/k_0 r\sigma_r)J_m(k_0\sigma_r r)\sin(k_0\sigma_z z+m\phi-\omega t+\varphi)\,\hat{\boldsymbol{r}}$$
$$+\dot{J}_m(k_0\sigma_r r)\cos(k_0\sigma_z z+m\phi-\omega t+\varphi)\,\hat{\boldsymbol{\phi}}\}$$
$$\times\big\{(\sigma_z/\sigma_r)\dot{J}_m(k_0\sigma_r r)\sin(k_0\sigma_z z+m\phi-\omega t)\,\hat{\boldsymbol{r}}$$
$$+(m\sigma_z/k_0 r\sigma_r^2)J_m(k_0\sigma_r r)\cos(k_0\sigma_z z+m\phi-\omega t)\,\hat{\boldsymbol{\phi}}$$
$$-J_m(k_0\sigma_r r)\cos(k_0\sigma_z z+m\phi-\omega t)\,\hat{\boldsymbol{z}}\}$$
$$= \mu_0|\chi|H_0^2(k_0/\sigma_r)$$
$$\times\{(m^2\sigma_z/k_0^2 r^2\sigma_r^3)J_m^2(k_0\sigma_r r)\sin(k_0\sigma_z z+m\phi-\omega t+\varphi)\cos(k_0\sigma_z z+m\phi-\omega t)\,\hat{\boldsymbol{z}}$$
$$-(\sigma_z/\sigma_r)\dot{J}_m^2(k_0\sigma_r r)\cos(k_0\sigma_z z+m\phi-\omega t+\varphi)\sin(k_0\sigma_z z+m\phi-\omega t)\,\hat{\boldsymbol{z}}$$
$$+(m/k_0 r\sigma_r)J_m^2(k_0\sigma_r r)\sin(k_0\sigma_z z+m\phi-\omega t+\varphi)\cos(k_0\sigma_z z+m\phi-\omega t)\,\hat{\boldsymbol{\phi}}$$
$$-J_m(k_0\sigma_r r)\dot{J}_m(k_0\sigma_r r)\cos(k_0\sigma_z z+m\phi-\omega t+\varphi)\cos(k_0\sigma_z z+m\phi-\omega t)\,\hat{\boldsymbol{r}}\,\}.$$

$$\boldsymbol{r}\times\langle(\partial\boldsymbol{P}/\partial t)\times\mu_0\boldsymbol{H}\rangle = ½(m|\chi|\mu_0 H_0^2/\sigma_r^2)J_m^2(k_0\sigma_r r)\sin\varphi\,\hat{\boldsymbol{z}}. \quad (40)$$

$$\boldsymbol{P}\times\boldsymbol{E} = \mu_0|\chi|H_0^2(m/k_0 r\sigma_r^3)J_m(k_0\sigma_r r)\dot{J}_m(k_0\sigma_r r)\{\sin(k_0\sigma_z z+m\phi-\omega t+\varphi)\cos(k_0\sigma_z z+m\phi-\omega t)$$
$$-\cos(k_0\sigma_z z+m\phi-\omega t+\varphi)\sin(k_0\sigma_z z+m\phi-\omega t)\}\hat{\boldsymbol{z}}.$$

$$\boldsymbol{P}\times\boldsymbol{E} = (m|\chi|\mu_0 H_0^2/k_0 r\sigma_r^3)J_m(k_0\sigma_r r)\dot{J}_m(k_0\sigma_r r)\sin\varphi\,\hat{\boldsymbol{z}}. \quad (41)$$

Upon combining the above expressions, we find the time-averaged force-density along the $z$-axis, $\langle F_z(\boldsymbol{r},t)\rangle$, and the time-averaged torque-density, $\langle\boldsymbol{T}(\boldsymbol{r},t)\rangle$, experienced by the test-particle to be

$$\langle F_z(\boldsymbol{r},t)\rangle = (|\chi|\mu_0 H_0^2\,\omega\sigma_z/2\sigma_r^2 c)\big[(m/k_0 r\sigma_r)^2 J_m^2(k_0\sigma_r r)+\dot{J}_m^2(k_0\sigma_r r)\big]\sin\varphi. \quad (42)$$

$$\boldsymbol{P}\times\boldsymbol{E}+\boldsymbol{r}\times\langle(\partial\boldsymbol{P}/\partial t)\times\mu_0\boldsymbol{H}\rangle+\boldsymbol{r}\times\langle(\boldsymbol{P}\cdot\boldsymbol{\nabla})\boldsymbol{E}\rangle = \mu_0|\chi|H_0^2\{(m/k_0 r\sigma_r^3)J_m(k_0\sigma_r r)\dot{J}_m(k_0\sigma_r r)\sin\varphi$$
$$+½(m/\sigma_r^2)J_m^2(k_0\sigma_r r)\sin\varphi$$
$$-½(m/\sigma_r^2)[1-(m/k_0\sigma_r r)^2]J_m^2(k_0\sigma_r r)\sin\varphi$$
$$-(m/k_0 r\sigma_r^3)J_m(k_0\sigma_r r)\dot{J}_m(k_0\sigma_r r)\sin\varphi$$
$$+½(m/\sigma_r^2)\dot{J}_m^2(k_0\sigma_r r)\sin\varphi\}.$$

$$\langle \boldsymbol{T}(\boldsymbol{r},t)\rangle = (m|\chi|\mu_0 H_0^2/2\sigma_r^2)[(m/k_0 r\sigma_r)^2 J_m^2(k_0\sigma_r r) + \dot{J}_m^2(k_0\sigma_r r)]\sin\varphi\,\hat{\boldsymbol{z}}. \quad (43)$$

Since force is the rate of acquisition of mechanical linear momentum by the particle (i.e., $\boldsymbol{F} = \mathrm{d}\boldsymbol{\wp}/\mathrm{d}t$), a comparison of Eq.(42) with Eq.(38) reveals that the linear momentum along the $z$-axis picked up by the particle is $\sigma_z/c$ times the absorbed energy, consistent with the quantum-optical picture of photons having energy $\hbar\omega$ and linear momentum $\hbar\omega/c$. The factor $\sigma_z$ in the above expression indicates that only the fraction $k_z/k_0$ of the photon energy propagates along the $z$-axis; the remainder circulates around the $z$-axis and gives rise to angular momentum.

Given that torque is the rate of acquisition of mechanical angular momentum by the particle (i.e., $\boldsymbol{T} = \mathrm{d}\boldsymbol{L}/\mathrm{d}t$), a comparison of Eq.(43) with Eq.(38) reveals that the total angular momentum picked up by the particle is $m/\omega$ times the absorbed energy, consistent with the quantum-optical picture of photons having energy $\hbar\omega$ and orbital angular momentum $m\hbar$.

Note that the fraction of torque that is due to the term $\boldsymbol{P}\times\boldsymbol{E}$ given by Eq.(41) integrates to zero across the entire space, indicating that, upon spatial averaging, no net spin angular momentum is transferred to the test-particle. This observation is, of course, consistent with the fact that the $m^{\text{th}}$ order mode under consideration is a pure TE polarized mode. The same conclusion would apply to a pure TM mode. Only a superposition of TE and TM cylindrical harmonics with identical $m$, $\omega$, $\sigma_r$, and $\sigma_z$ can produce a beam that could be considered akin to a circularly polarized plane-wave, in which case the beam (upon spatial averaging) will transfer a net spin (as well as orbital) angular momentum to the test-particle.

**8. Concluding Remarks**. Assuming that a small polarizable test-particle is placed within an EM field, we have related the force and torque exerted on the particle to the energy that is taken away from the local field. The conservation laws of energy and momentum were then invoked to relate the linear and angular momenta of the EM field to its energy. Along the way, we showed the differing effects on the test-particle of the field's spin and orbital angular momenta.

The subject of optical angular momentum is vast, and many excellent monographs, reviews, and research papers have been devoted to this topic over the last several decades. The goal of the present paper has not been to break new ground, nor to report new discoveries. Rather, it has been our objective to bring to the reader's attention some of the basic properties of optical angular momentum that can be described in the context of simple examples using rigorous physical arguments rooted in the classical theory of electrodynamics. The mathematical operations needed to describe the physical phenomena are elementary for the most part, although some degree of familiarity with Bessel functions and vector algebra is necessary for a thorough understanding of the subject. It is hoped that our rigorous discussion of the principles using Maxwell's macroscopic equations and Poynting's theorem in conjunction with a consistent application of the expressions of electromagnetic force, torque, momentum, and angular momentum, has been helpful in illuminating some of the obscure and perhaps less-well-understood aspects of the theory of optical angular momentum. We also hope that the analysis has succeeded in demonstrating the consistency of the classical theory with the principles of conservation of energy and momentum, as well as revealing certain similarities between the classical and quantum-optical properties of angular momentum.